\documentclass[sigconf]{acmart}

\AtBeginDocument{%
  }

\usepackage{cleveref}
\usepackage{lipsum}
\usepackage{enumitem} 
\usepackage{svg}
\usepackage{graphicx}
\usepackage{graphicx}
\usepackage{epstopdf}
\usepackage{float}
\usepackage{subcaption}
\usepackage{colortbl}
\usepackage{soul}

\setcopyright{acmlicensed}
\copyrightyear{2025}
\acmYear{2025}
\acmConference[UbiComp Companion '25] {Companion of the 2025 ACM International Joint Conference on Pervasive and Ubiquitous Computing}{October 12--16, 2025}{Espoo, Finland.}
\acmISBN{979-8-4007-1477-1/25/10}
\acmDOI{10.1145/3714394.3756289}

\begin{document}

\title{Can Large Language Models Identify Materials from Radar Signals?}

\author{Jiangyou Zhu}
\orcid{0000-0003-1512-6784}
\affiliation{
  \institution{The Chinese University of Hong Kong}
  \country{Hong Kong SAR, China}}
\email{zj124@ie.cuhk.edu.hk}

\author{Hongyu Deng}
\orcid{0009-0003-5935-9050}
\affiliation{
  \institution{The Chinese University of Hong Kong}
  \country{Hong Kong SAR, China}}
\email{dh021@ie.cuhk.edu.hk}

\author{He Chen}
\orcid{0000-0001-8886-9680}
\authornote{He Chen is the corresponding author.}
\affiliation{
  \institution{The Chinese University of Hong Kong}
  \country{Hong Kong SAR, China}}
\email{he.chen@ie.cuhk.edu.hk}

\renewcommand{\shortauthors}{Jiangyou Zhu, Hongyu Deng, \& He Chen}

\begin{teaserfigure}
  \centering
  \includegraphics[width=\textwidth]{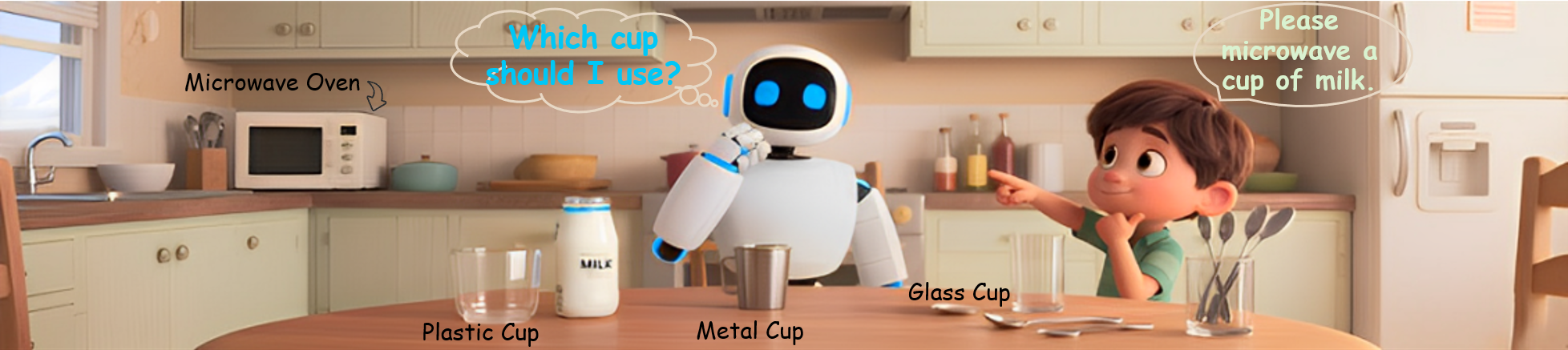}
  \caption{Home robots often need to distinguish the material composition of visually similar objects during everyday tasks. This figure illustrates such a scenario, where the robot needs to identify a glass cup to safely microwave a cup of milk, a task made challenging by its visual similarity to a plastic cup, which may not be microwave-safe. Consequently, additional sensing modalities or methods are necessary to achieve accurate material identification. This image is generated by the Doubao AI model from a sketch input, with additional text overlaid on the generated image.}
  \label{fig1}
\end{teaserfigure}

\begin{abstract}
Accurately identifying the material composition of objects is a critical capability for AI robots powered by large language models (LLMs) to perform context-aware manipulation. Radar technologies (e.g., millimeter-wave radar) offer a promising sensing modality for material recognition tasks, providing robustness to lighting conditions and high spatial resolution. When combined with deep learning, radar technologies have demonstrated strong potential in identifying the material composition of various objects. However, existing radar-based solutions are often constrained to closed-set object categories and typically require task-specific data collection to train deep learning models, largely limiting their practical applicability. This raises an important question: \textit{Can we leverage the powerful reasoning capabilities of pre-trained LLMs to directly infer material composition from raw radar signals}? Answering this question is non-trivial due to the inherent redundancy of radar signals and the fact that pre-trained LLMs have no prior exposure to raw radar data during training. To address this, we introduce LLMaterial, the first study to investigate the feasibility of using LLMs to identify materials directly from radar signals. First, we introduce a physics-informed signal processing pipeline that distills high-redundancy radar raw data into a set of compact intermediate parameters that encapsulate the material’s intrinsic characteristics. Second, we adopt a retrieval-augmented generation (RAG) strategy to provide the LLM with domain-specific knowledge, enabling it to interpret and reason over the extracted intermediate parameters. Leveraging this integration, the LLM is empowered to perform step-by-step reasoning on the condensed radar features, achieving open-set material recognition directly from raw radar signals. Preliminary results show that LLMaterial can effectively distinguish among a variety of common materials, highlighting its strong potential for real-world material identification applications.
\end{abstract}

\ccsdesc[500]{Human-centered computing~Ubiquitous and mobile computing systems and tools}

\keywords{Material identification, radar sensing, large language model, retrieval-augmented generation.}

\maketitle

\section{Introduction}
Object recognition is a fundamental capability for autonomous systems, underpinning a wide range of applications in home robotics, assistive technologies, and industrial automation~\cite{zou2023object}. 
In recent years, the application of AI to vision-based object recognition has made substantial strides. In particular, the rise of large language models (LLMs) has led to remarkable results not only in object identification but also in recognizing object attributes such as material composition~\cite{yu2024towards, gouidis2024llm, zhao2024octopus}. This progress is largely attributed to a key strength of LLMs: their deep semantic understanding, developed through large-scale pretraining on diverse and extensive datasets. However, our real-world experiments reveal a critical limitation: LLMs relying solely on visual inputs often fail to distinguish between visually similar objects, which can lead to unsafe decisions. For instance, a robot may be unable to differentiate between a glass and a plastic cup, an error that poses a safety risk when heating liquids, as only glass is microwave-safe (see~\Cref{fig1}). This challenge has also been highlighted in prior work~\cite{cao2025objvariantensemble}, emphasizing the need for more robust, multimodal object recognition systems that extend beyond vision alone.

\begin{figure*}
  \includegraphics[width=\textwidth]{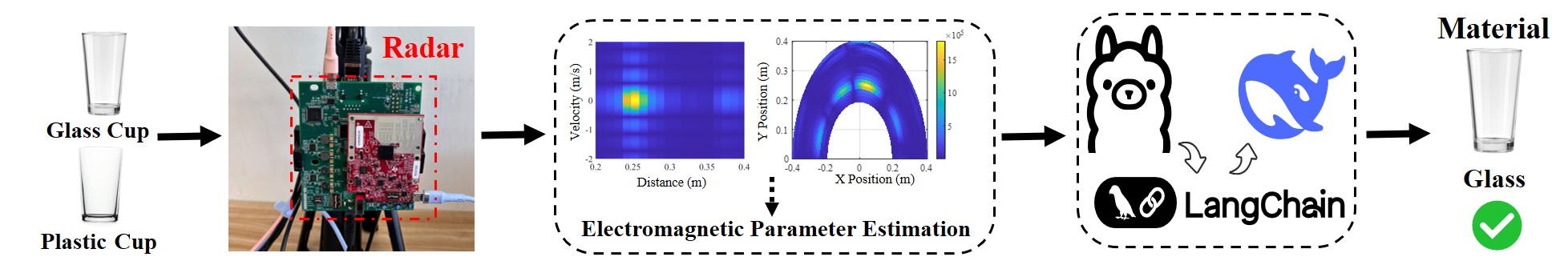}
  \caption{Overview of LLMaterial.}
  \Description{For visually similar objects, LLMaterial employs radar to detect targets and analyzes the received signals to extract electromagnetic properties, which are subsequently processed by an LLM via the Ollama framework, utilizing LangChain for enhanced material recognition.}
  \label{fig2}
\end{figure*}

As an emerging sensing modality, radar is increasingly recognized for its noticeable potential in material recognition. Specifically, RadarCat \cite{yeo2016radarcat} pioneered this direction by using time-domain analysis of near-field millimeter-wave (mmWave) signals and machine learning to classify materials based on their unique reflective signatures. 
mSense \cite{wu2020msense} quantitatively characterizes the material’s reflectivity using the material reflection feature (MRF), and then determines the material type by finding the best match against a pre-trained database storing the MRFs of different materials. 
CRFUSION \cite{xiao2025crfusion} combines camera and radar modalities to simultaneously identify object categories and materials, using the energy reflection factor (ERF) as a key feature within a modality fusion network. However, a common limitation of these approaches is that they often rely on extensive data collection to train end-to-end material identification models. However, their effectiveness is typically confined to a narrow range of materials, which noticeably limits their practical utility in real-world applications.
Given the profound comprehension capabilities of LLMs, a promising direction is to leverage them for interpreting radar signals in material recognition tasks. This leads us to our research question: \textit{Can LLMs infer an object's electromagnetic properties and accurately identify its material from radar signals in an open-set setting?}

Answering this question is nontrivial, as it involves overcoming challenges stemming from both the limitations of the LLM’s prior knowledge and the complex nature of raw radar data.
First, because radar raw data is rarely represented in the pretraining datasets of LLMs, these models lack the inherent ability to directly interpret the semantic content embedded in radar signals. This necessitates a dedicated preprocessing step to translate raw radar inputs into a representation that aligns with the LLM’s modality and semantic understanding. 
In addition, the high dimensionality and inherent redundancy of raw radar data present significant challenges. Much of the signal may be unrelated to the material’s intrinsic properties, introducing noise that can obscure relevant information. As a result, a key challenge lies in guiding the LLM to focus selectively on material-relevant features within the radar signal.

This paper introduces LLMaterial, the first system that can leverage pre-trained LLMs to directly identify object materials from raw radar signals. Our contributions in developing LLMaterial are threefold:
\begin{itemize}[left=0pt, itemsep=0pt, topsep=0pt]
    \item We develop a physics-informed signal processing pipeline designed to distill a material's intrinsic dielectric constant from raw radar signals. This multi-stage process systematically reduces data redundancy, transforming the high-dimensional signal into a set of compact intermediate parameters of the material’s intrinsic properties. This approach provides more tractable and interpretable input for subsequent LLM reasoning.
    \item We incorporate a retrieval-augmented generation (RAG) framework to equip the LLM with domain-specific knowledge, enabling it to construct a chain-of-thought (CoT) that reasons about the interplay of physical properties. For example, when presented with an object made of a specialized material exhibiting a high dielectric constant but a low radar cross section (RCS), a conventional classifier may struggle to resolve the ambiguity and produce an accurate prediction. In contrast, LLMaterial is designed to reason through such scenarios step by step, inferring that this combination of properties likely corresponds to a radar-absorbent material. Rather than relying on a direct feature-to-label mapping, LLMaterial can interpret underlying physical characteristics through semantic reasoning.    
    \item We perform an initial validation to assess the practical feasibility of LLMaterial. The results indicate that LLMaterial can reliably identify four representative materials, including metal, ceramic, glass, and plastic, demonstrating its effectiveness in real-world material recognition tasks. This validation underscores the promise of our approach and provides a strong foundation for future research toward more intelligent and context-aware assistive technologies.

\end{itemize}

We note that recent studies have explored the use of LLMs to interpret data from other non-visual sensors, such as inertial measurement units (IMUs)~\cite{leng2024imugpt, wang2025ego4o} and ultra-wideband (UWB) sensors~\cite{yang2025nlos}.

\section{System Design}

In this section, we introduce the architecture of LLMaterial, designed to perform robust material identification through a new integration of radar sensing and LLM-based reasoning. In our work, we used an mmWave radar sensor (IWR6843ISK) manufactured by Texas Instruments (Dallas, TX, USA). The overall workflow of LLMaterial is depicted in \Cref{fig2}. For visually similar objects, LLMaterial first acquires the raw target echo signal using an mmWave radar. Next, this raw radar data is preprocessed, followed by performing electromagnetic parameter estimation to extract key parameters. These parameters are then fed into an LLM-based system augmented with specialized domain knowledge. Leveraging RAG, the model then generates a CoT process to reason about all electromagnetic properties, ultimately leading to robust material identification. 

\subsection{Radar Signal Processing}
\begin{figure}[!t]
\centering
\begin{minipage}{0.48\linewidth}
\centering
\includegraphics[width=\linewidth]{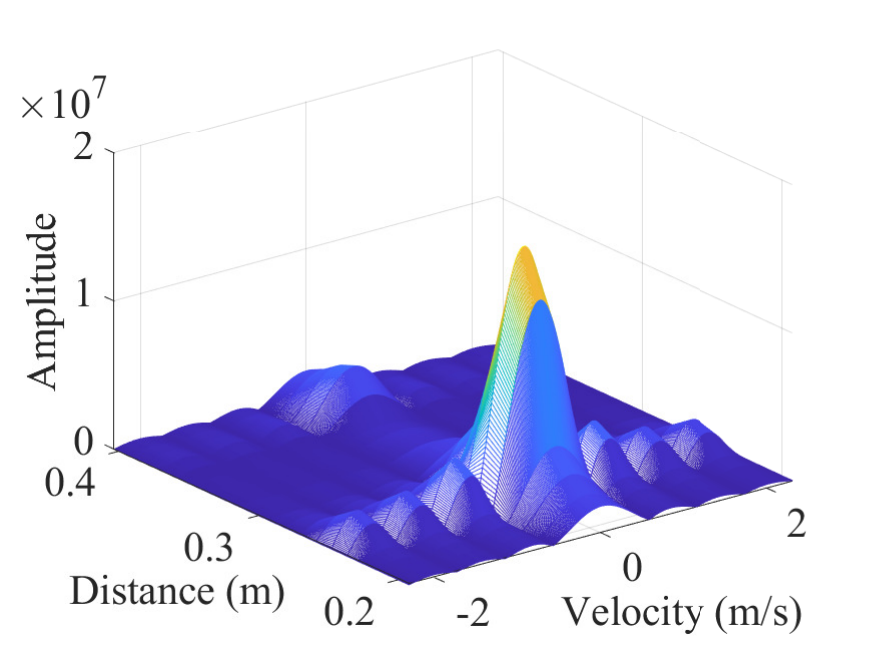}
\end{minipage}
\hfill
\begin{minipage}{0.48\linewidth}
\centering
\includegraphics[width=\linewidth]{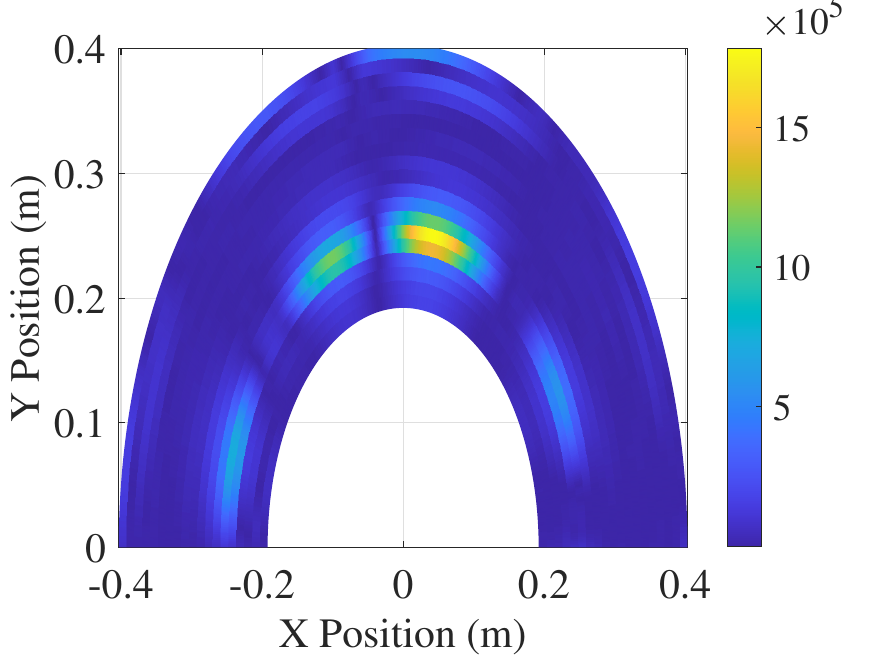}
\end{minipage}
\caption{(Left) RD map; (Right) RA map.}
\Description{The RD and RA maps facilitate the rapid localization of objects, enabling the extraction of signals from the corresponding positions for subsequent analysis.}
\label{fig3}
\end{figure}

This part details our signal preprocessing pipeline. We first perform multichannel accumulation on the 3D radar echoes to yield a high-resolution range-Doppler (RD) map. Subsequently, the range-angle (RA) map is derived by processing the data in the slow-time dimension. The combination of these maps provides the target's distance ($R$), velocity ($V$), and angle ($\theta$) relative to the radar, which are crucial for its accurate localization. Representative RD and RA maps are depicted in \Cref{fig3}.

\begin{figure}[!t]
\centering
\begin{minipage}{0.48\linewidth}
\centering
\includegraphics[width=\linewidth]{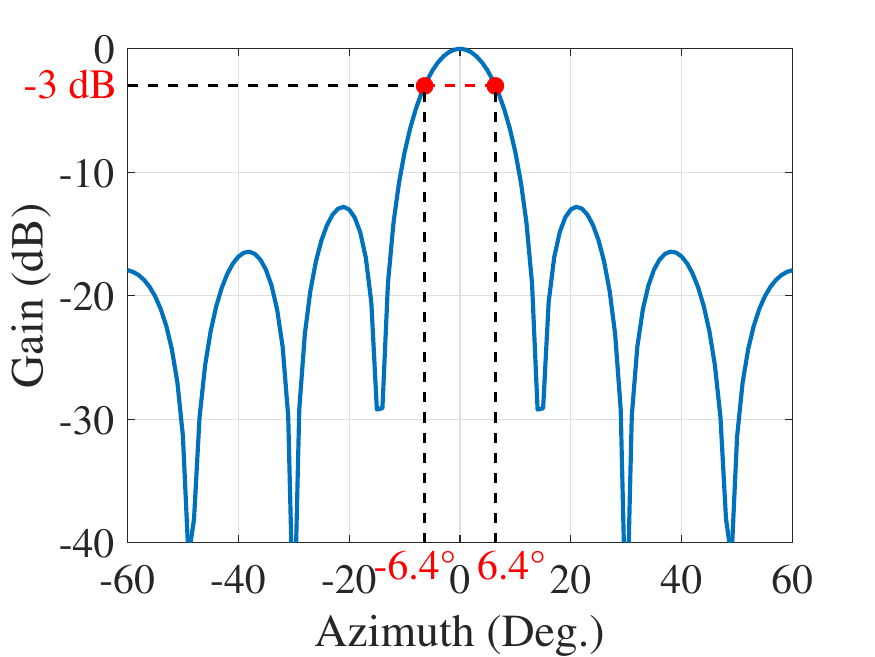}
\end{minipage}
\hfill
\begin{minipage}{0.48\linewidth}
\centering
\includegraphics[width=\linewidth]{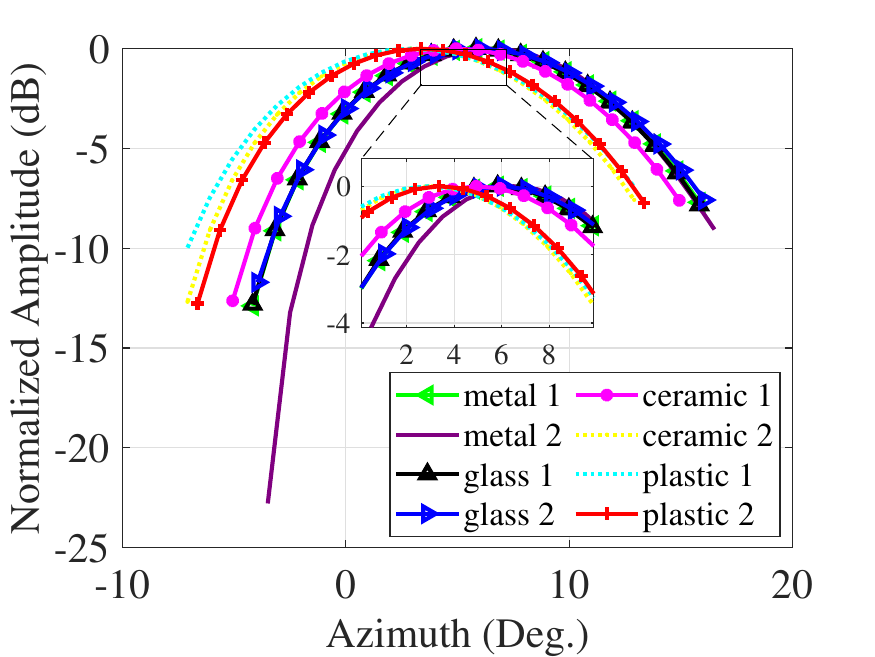}
\end{minipage}
\caption{(Left) Beam Pattern of the Antenna Array; (Right) Beam Direction After Beamforming.}
\Description{The array aperture results in a wider beam width; however, utilizing beamforming techniques can get more precise angular information about the target.}
\label{fig4}
\end{figure}

To enhance the precision of target angle estimation, we employ beamforming techniques to accurately resolve the target's direction of arrival (DoA).  This step is crucial, as the antenna array's native resolution is limited by its physical aperture; a smaller aperture results in a wider half-power beamwidth (HPBW). As shown in \Cref{fig4}, the HPBW of our mmWave radar array was measured to be $\ 12.8^\circ$, and the target angle after beamforming can achieve better accuracy. 

The aforementioned steps constitute our signal preprocessing stage. Once a target is detected and its signal is extracted, the subsequent analysis focuses on electromagnetic parameter estimation, as shown in \Cref{fig5}. First, for a target that meets far-field conditions, the received echo power is determined by the classical radar equation, expressed as:
\begin{equation}
P_r = P_t G_t G_r \frac{\lambda^2 \sigma}{(4\pi)^3 R^4},
\label{eq1}
\end{equation}
where $P_r$ is the received power, $P_t$ is the transmitted power, $G_t$ and $G_r$ are the gains of the transmitting and receiving antennas, respectively, $\lambda$ is the signal wavelength, $\sigma$ is the target's RCS, and $R$ is the distance of the target relative to the mmWave radar. Based on the expression of $P_r$, the signal-to-noise ratio (SNR) of the target echo signal can be expressed as:
\begin{equation}
SNR = \frac{P_r}{P_n} =  \frac{P_t G_t G_r \lambda^2 \sigma}{(4\pi)^3 k T_n B R^4} = K \frac{\sigma}{R^4},
\label{eq2}
\end{equation}
where $P_n$ is the total noise power of the receiver. For thermal noise, $P_n = k T_n B$, where $k$ is the Boltzmann constant, $T_n$ is the effective system noise temperature, and $B$ is the bandwidth. The constant $K$ combines all system-specific parameters, highlighting that the SNR is primarily dependent on the target's RCS and its distance. Constant $K$ can be expressed as:
\begin{figure}[t]  \includegraphics[width=0.48\textwidth]{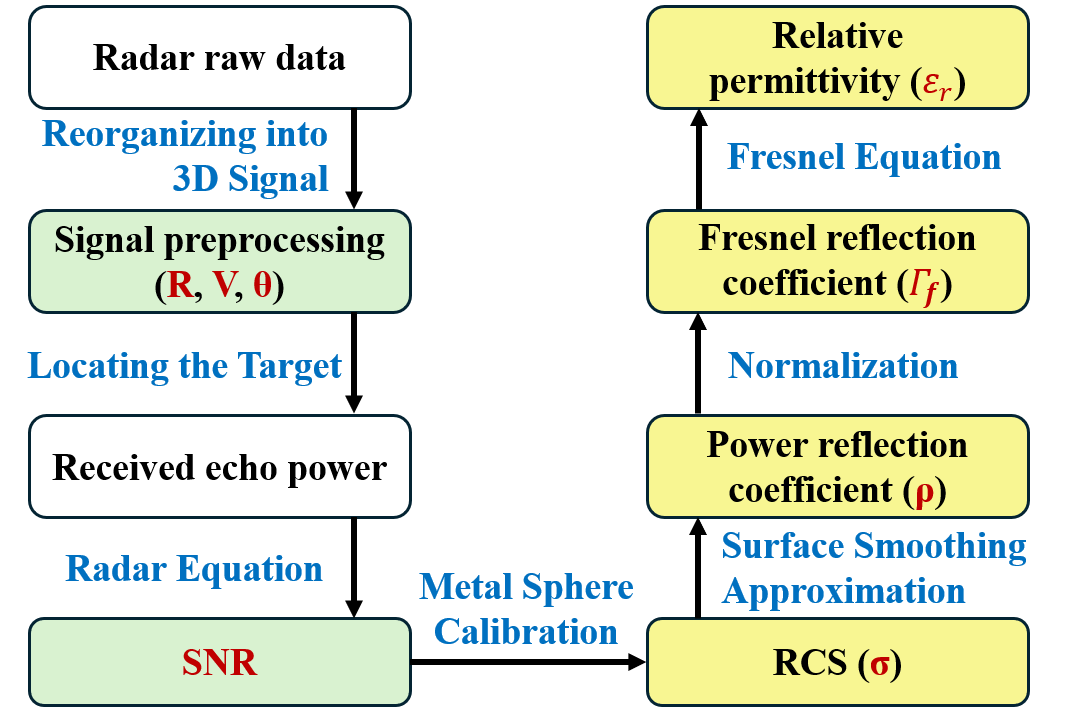}
  \caption{Workflow of radar signal processing.}
  \Description{This flowchart illustrates the process of analyzing radar raw data. It begins with reorganizing the radar raw data into a 3D signal format. The next step involves signal preprocessing, which includes locating the target and extracting relevant parameters ($R$, $V$, $θ$). The received echo power is then utilized in conjunction with the radar equation to compute the SNR. On the right side, through metal sphere calibration, we get the RCS ($\sigma$), followed by calculating the power reflection coefficient ($\rho$) and the Fresnel reflection coefficient ($\Gamma_f$). Finally, the relative permittivity ($\varepsilon_r$) is derived using the Fresnel equation.}
  \label{fig5}
\end{figure}

\begin{equation}
K = \frac{P_t G_t G_r \lambda^2}{(4\pi)^3 k T_n B}.
\label{eq3}
\end{equation}

To determine the system constant $K$, we performed a calibration using a metal sphere ($d = 63$\, mm) as a standard target. For a sphere with a diameter significantly exceeding the signal wavelength ($\lambda = 5$\, mm), the sphere's RCS in the optical scattering regime equals its physical cross-sectional area, which can be calculated as:
\begin{equation}
    \sigma_c = \pi \left( \frac{d}{2} \right)^2 \approx 0.0031\,\text{m}^2.
    \label{eq4}
\end{equation}

After measuring the SNR of the sphere, which serves as a known reference, we computed $K$ by inversion of \Cref{eq2}. This calibrated constant enables the direct calculation of an unknown target's RCS, which is given by:
\begin{equation}
\sigma = SNR \frac{R^4}{K}.
\label{eq5}
\end{equation}

However, the RCS itself is a composite parameter, integrating the intrinsic electromagnetic properties of the target's material with its overall geometric shape. To disentangle these attributes and analyze the material characteristics more directly, we relate the RCS to the backscattering coefficient. In the field of radar, the backscattering coefficient, denoted as $\sigma_0$, is defined as the average RCS per unit area, where $A_c$ denotes the area on the ground associated with that object. The RCS of a specific object can be expressed as:
\begin{equation}
\sigma = \sigma_0 A_c.
\label{eq6}
\end{equation}

\begin{figure}[!t]
\centering 
\begin{minipage}[t]{0.46\columnwidth}
    \centering
    \includegraphics[width=\linewidth]{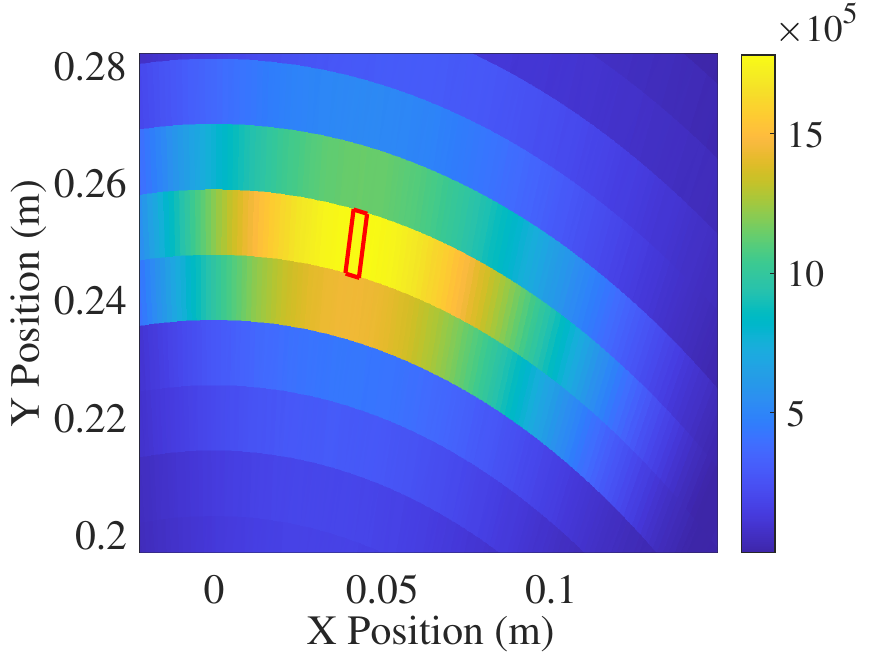} 
    \caption{PRCA (red box).}
    \label{fig6}
    \Description{The PRCA is defined by selecting the largest unit cell with the highest peak value within the target area. This approach approximates the local surface as smooth, allowing for the determination of reflection capability per unit area, which subsequently characterizes the material properties.}
\end{minipage}%
\begin{minipage}[t]{0.46\columnwidth}
    \centering
    \includegraphics[width=0.9\linewidth]{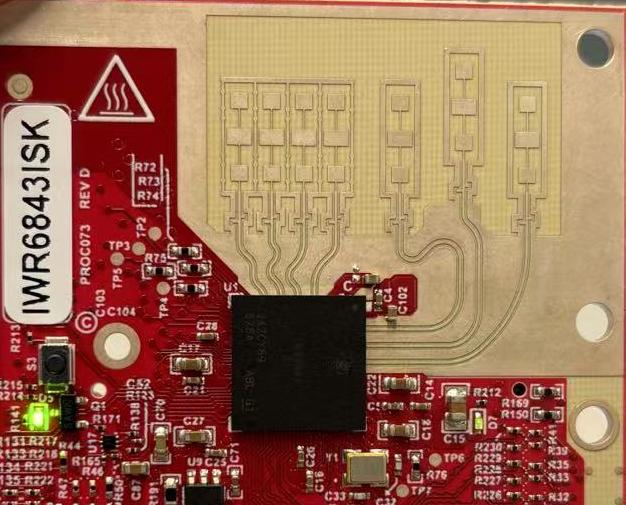} 
    \caption{Antenna array.}
    \label{fig7}
    \Description{We used an mmWave radar sensor (IWR6843ISK) produced by Texas Instruments, with the antenna array positioned horizontally, resulting in vertical polarization.}
\end{minipage}
\end{figure}

Inspired by this principle, we shift our focus from the total RCS to the dominant reflection center that generates the signal peak. We define a power reflection coefficient $\rho$ to represent the strong reflection signal originating from a small but highly reflective region, which we term as peak reflection cell area (PRCA) $A_r$. The extracted PRCA is shown in \Cref{fig6}. By approximating PRCA as a local specular reflector, $\rho$ becomes analogous to $\sigma_0$. Therefore, RCS can be further expressed as:
\begin{equation}
\sigma = \rho A_r.
\label{eq7}
\end{equation}

Theoretically, the power reflection coefficient $\rho$ is described by the Fresnel equations, which define it as a function of the relative permittivity ($\varepsilon_r$) of the material, the polarization of the wave, and the angle of incidence. For the positive $\rho$ that we calculate, we perform a normalization to obtain the Fresnel reflection coefficient $\Gamma_f$ for vertical polarization. This step is justified by our core assumptions: specular reflection from a locally smooth surface combined with the known vertical polarization of our mmWave radar, as shown in \Cref{fig7} (as solving for $\varepsilon_r$ with horizontal polarization is mathematically intractable, resulting in a high-order polynomial, and our approach is consistent with commercial radars, which are commonly vertically polarized, thereby focusing our analysis to the vertical polarization case). Consequently, $\Gamma_f$ is modeled as:
\begin{equation}
\Gamma_f = \frac{\varepsilon_r \cos \theta - \sqrt{\varepsilon_r - \sin^2 \theta}}{\varepsilon_r \cos \theta + \sqrt{\varepsilon_r - \sin^2 \theta}},
\label{eq8}
\end{equation}
where $\varepsilon_r = \varepsilon' - j \varepsilon''$, $\varepsilon'$ is the real part, and $\varepsilon''$ is the imaginary part.

We define the variables as follows:
\begin{equation}
A = \varepsilon_r \cos \theta, \quad B = \sqrt{\varepsilon_r - \sin^2 \theta}.
\label{eq9}
\end{equation}

Then, we get:
\begin{equation}
\Gamma_f = \frac{A - B}{A + B}.
\label{eq10}
\end{equation}

We can further rewrite it as:
\begin{equation}
\Gamma_f (A + B) = A - B.
\label{eq11}
\end{equation}

Next, we expand and rearrange to obtain:
\begin{equation}
\begin{aligned}
\Gamma_f A + \Gamma_f B &= A - B, \\
\Gamma_f A - A &= -\Gamma_f B - B, \\
A(\Gamma_f - 1) &= -B(\Gamma_f + 1).
\label{eq12}
\end{aligned}
\end{equation}

Substituting A and B back into \Cref{eq12}, we have:
\begin{equation}
\varepsilon_r \cos \theta (\Gamma_f - 1) = -\sqrt{\varepsilon_r - \sin^2 \theta} (\Gamma_f + 1).
\label{eq13}
\end{equation}

To proceed, we square both sides, expand and rearrange, and move all terms to the left side, getting:
\begin{equation}
\varepsilon_r^2 \cos^2 \theta (\Gamma_f^2 - 2\Gamma_f + 1) - \varepsilon_r (\Gamma_f + 1)^2 + \sin^2 \theta (\Gamma_f + 1)^2 = 0.
\label{eq14}
\end{equation}

We treat \Cref{eq14} as a quadratic equation in terms of $\varepsilon_r$, and then we can solve it using the quadratic formula. Due to space constraints, we do not provide an extensive derivation here. The final relationship between $\varepsilon_r$, $\Gamma_f$ and $\theta$ is given by:
\begin{equation}
\varepsilon_r = \frac{(\Gamma_f + 1)^2 \left[ 1 \pm \sqrt{1 - \left( \frac{\sin 2\theta (\Gamma_f - 1)}{\Gamma_f + 1} \right)^2} \right]}{2 \cos^2 \theta [\Gamma_f - 1]^2}.
\label{eq15}
\end{equation}

According to the principles of physics, for $\varepsilon_r$, which is a complex number, we choose the positive root. Specifically, when \(\theta = 0\) (i.e., the target is located directly in front of the radar), we can get:
\begin{equation}
\varepsilon_r = \left( \frac{1 + \Gamma_f}{1 - \Gamma_f} \right)^2.
\label{eq16}
\end{equation}

In summary, we have systematically traced the signal processing workflow from initial raw signal to SNR, then through the analysis of RCS ($\sigma$), power reflection coefficient ($\rho$), and Fresnel reflection coefficient ($\Gamma_f$), ultimately leading to the extraction of the target's fundamental material property: the relative permittivity ($\varepsilon_r$). This signal processing pipeline facilitates the extraction of essential physical characteristics of an object directly from raw radar signals.

\subsection{RAG-Enhanced LLM}
\begin{figure*}[!t]
  \includegraphics[width=0.47\textwidth]{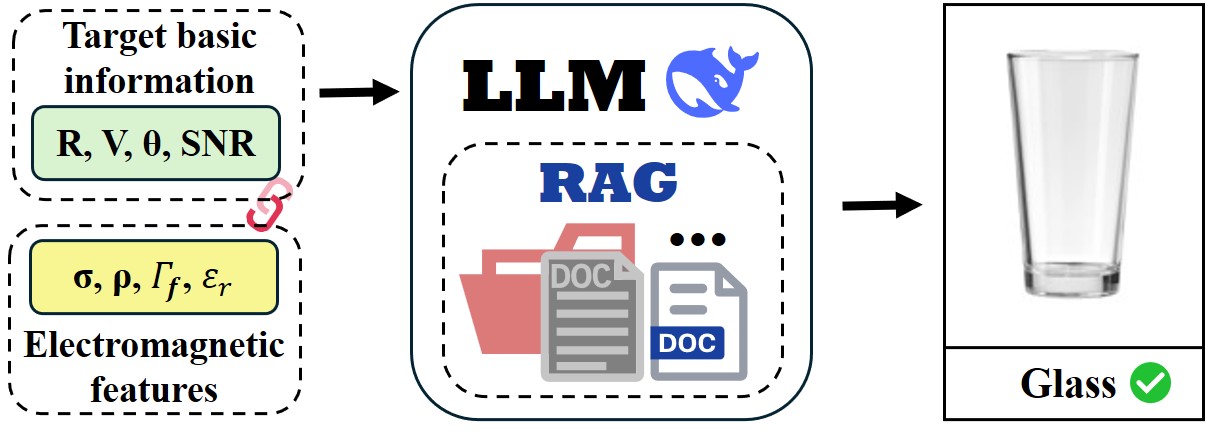}
  \caption{Workflow of the RAG-enhanced LLM for Radar-based material identification system.}
  \Description{The model processes user queries using eight key radar parameters, including the target's distance, velocity, angle, and SNR, along with four electromagnetic features: RCS, power reflection coefficient, Fresnel reflection coefficient, and relative permittivity. These parameters are input into the RAG system, which integrates an LLM with a domain-specific knowledge base. The process begins by creating a Doc Subset through document segmentation into chunks, which are then embedded and indexed in a Data Storage (vector database). During a query, a semantic search retrieves the Top-K most relevant chunks, which, along with engineered prompts, condition the LLM to generate accurate responses. This repeatable process fosters a specialized domain memory, enhancing material identification for assistive systems.}
  \label{fig8}
\end{figure*}

In this section, we present our LLM-based framework for material inference. As illustrated in \Cref{fig8}, the LLM interprets eight parameters extracted from the radar signal processing pipeline to predict the material composition of the target object. These include the target's basic information ($R$, $V$, $\theta$) and SNR, along with four key electromagnetic features: $\sigma$, $\rho$, $\Gamma_f$, and $\varepsilon_r$.

As shown in \Cref{fig8}, we apply the RAG technique to enhance the LLM's capability in material identification. The RAG process establishes a foundation for the LLM within a domain-specific knowledge base through a structured workflow. Initially, a doc subset is created by segmenting documents into chunks, which are then embedded and indexed in a specialized data storage (i.e., vector database). When the database is ready, LLMaterial performs a semantic search to retrieve the top-k chunks that are most relevant from this storage. These retrieved chunks, along with engineered prompts, are then used to inform the LLM, which generates an accurate response supported by the provided information. This process makes the model's reasoning transparent, as the output can be traced back to specific source documents. As a result, LLMaterial provides not only a correct recognition result but also a trustworthy and verifiable one.

\begin{figure*}[!t]
\centering
\begin{minipage}{\linewidth}
\centering
\includegraphics[width=0.87\textwidth]{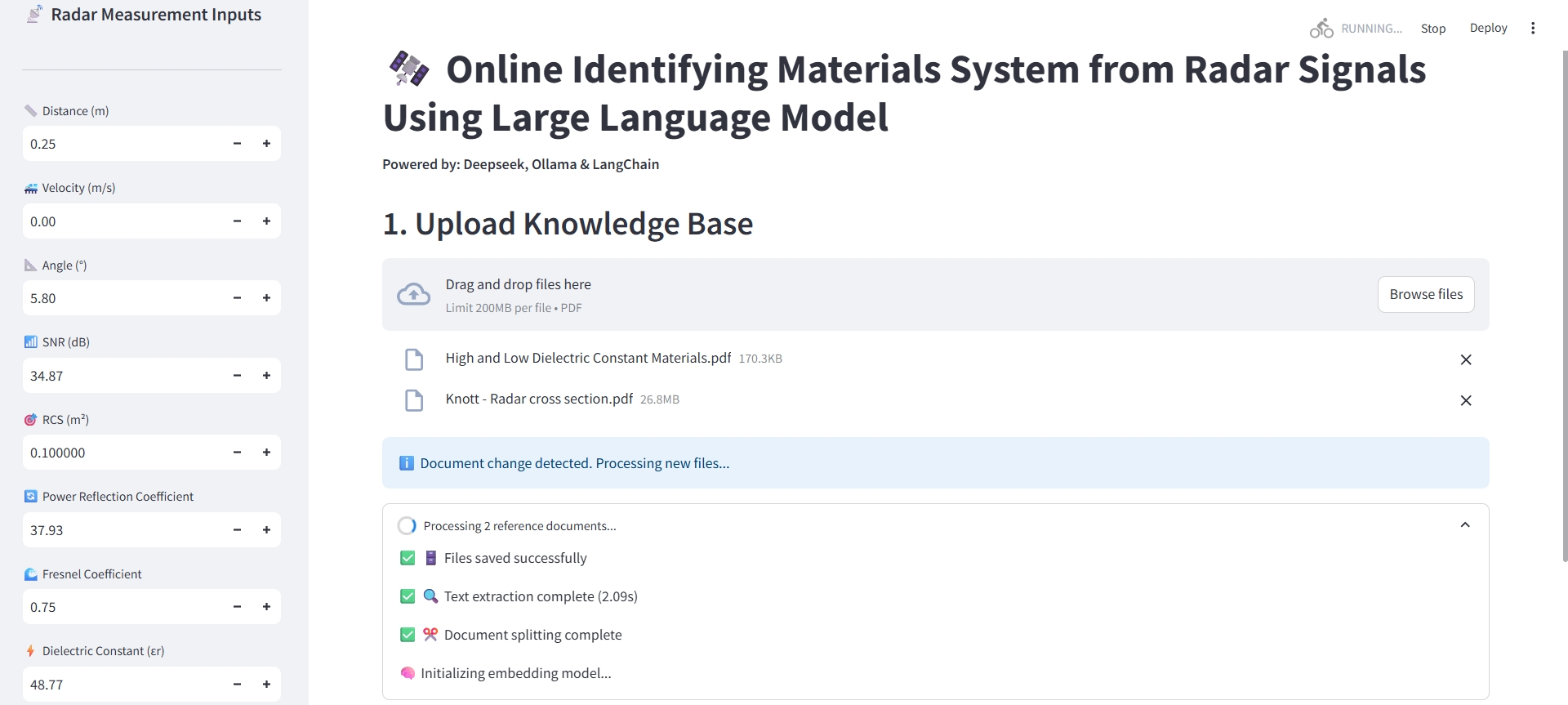}
\end{minipage}
\hfill
\begin{minipage}{\linewidth}
\centering
\includegraphics[width=0.87\textwidth]{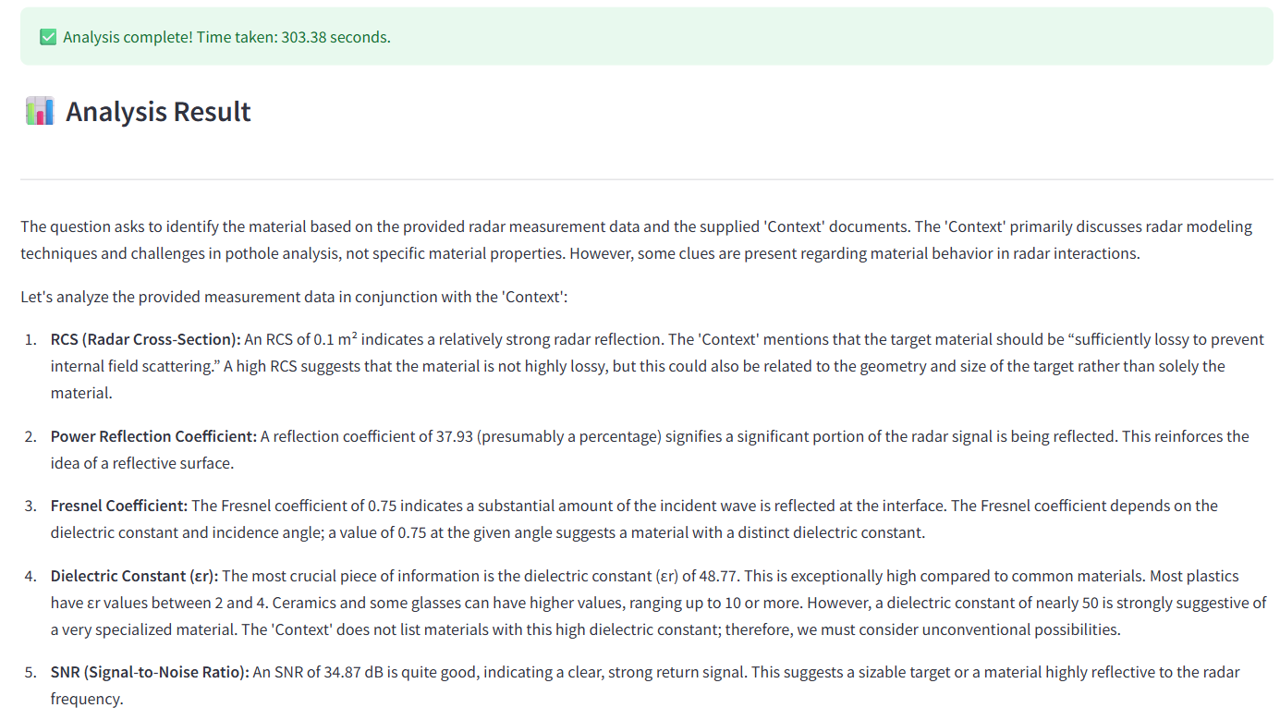}
\end{minipage}
\caption{Online material identification from radar signals using our LLM-based system. (Up) Step 1; (Down) Step 2.}
\Description{The user interface of the online identifying materials system from radar signals using LLM is designed for an efficient workflow, featuring two main panels. The left panel displays eight essential parameters derived from mmWave radar signal processing, which serve as quantitative inputs for the analysis. The right panel guides users through the process of uploading domain-specific documents to establish the knowledge base for the RAG system, offering a visual representation of the internal workflow for clarity. After successfully uploading the documents, users can initiate the analysis, allowing the system to process the input parameters and generate material identification results.}
\label{fig9}
\end{figure*}

As shown in \Cref{fig9}, the online identification system of our proposed LLMaterial is divided into two main panels:

\begin{itemize}[left=0pt, itemsep=0pt, topsep=0pt]
    \item \textbf{Left Panel (Radar Measurement Inputs):} This panel displays the eight key parameters derived from the mmWave radar signal processing workflow. These parameters serve as the quantitative input for the following analysis.
    \item \textbf{Right Panel (Online analysis of LLM):} This panel provides a guided two-step analytical process.\\
    (1) \textbf{Step 1: Upload Knowledge Base.} The user first uploads a set of domain-specific documents to serve as the knowledge base for the RAG system. Upon successful upload, the interface visualizes the internal RAG workflow, providing transparency into the retrieval and augmentation process. This allows users to monitor operations easily.\\
    (2) \textbf{Step 2: Start Analysis.} Once the knowledge base is loaded, the user can initiate the analysis. The system then processes the eight parameters, allowing the LLM to use RAG and CoT for comprehensive analysis and inference, ultimately generating the final material identification result of a real-world object.
\end{itemize}

\begin{figure*}[!t]
  \includegraphics[width=\textwidth]{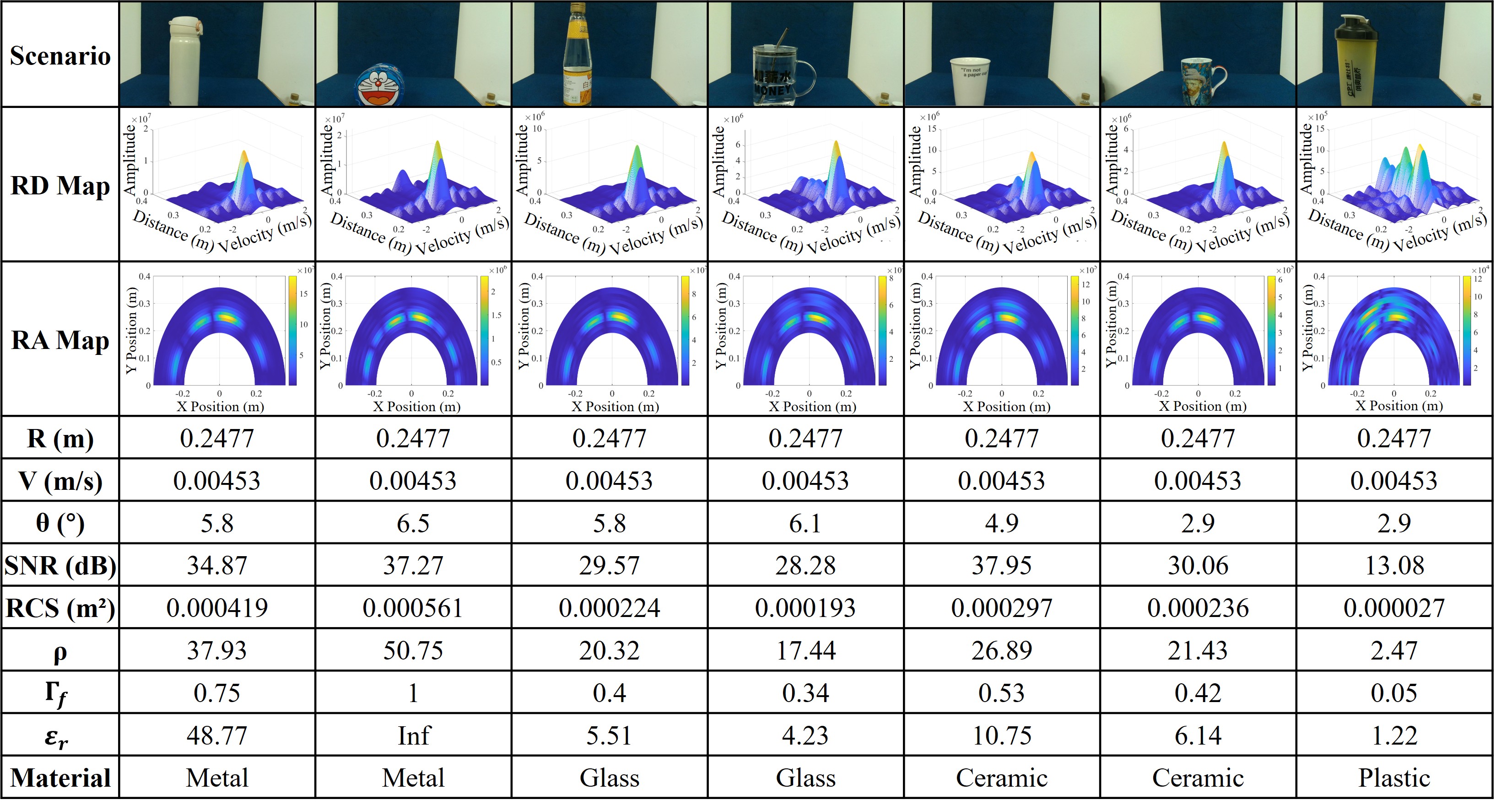}
  \caption{Summary of radar signal processing and recognition results: 7 common objects and 4 material types.}
  \Description{This figure summarizes the radar signal processing and recognition results for seven common objects and four material types. The first row corresponds to the identification scenarios, followed by the preprocessed RD and RA maps. Next, it presents the parameters of distance $R$, velocity $V$, angle $\theta$, and  SNR, along with four key electromagnetic features: RCS $\sigma$, power reflection coefficient $\rho$, Fresnel reflection coefficient $\Gamma_f$, and relative permittivity $\varepsilon_r$. Finally, the material identification results are displayed, clearly indicating a complete match with the actual materials.}
  \label{fig10}
\end{figure*}

Moreover, \Cref{fig9} shows the noticeable impact of RAG's knowledge base on the LLM's reasoning. As an initial effort to provide domain-specific context, our knowledge base was constructed from primary sources on key topics, such as RCS \cite{knott2004radar} and the relationship between dielectric constants and material properties \cite{singh1999high}. 

\section{Preliminary Experimental results and analysis}
\begin{table}
  \caption{Radar Parameter Settings.}
  \label{tab1}
  \begin{tabular}{ccl}
    \toprule
    Parameter&Value&Explanation\\
    \midrule
    $f_0$ & $60\,\mathrm{GHz}$ & Carrier Frequency\\
    $S$ & $66\,\mathrm{MHz}/\mu\mathrm{s}$ & Modulation Slope\\
    $B$ & $3.96\,\mathrm{GHz}$ & Bandwidth\\
    $f_s$ & $10\,\mathrm{MHz}$ & Sampling Rate\\
  \bottomrule
\end{tabular}
\end{table}
The parameters of mmWave radar are listed in \Cref{tab1}. Our signal processing pipeline was implemented in MATLAB R2023a. The 14-billion-parameter DeepSeek LLM (deepseek-r1:14b) was employed. All experiments were conducted on a Windows desktop equipped with a 2.60 GHz Intel Core i7-9750H CPU and 16 GB of RAM.

{\textbf{Experimental results}}: To evaluate LLMaterial's performance, we tested seven common objects made from four distinct material categories, with results summarized in \Cref{fig10}. The preliminary findings show that the electromagnetic parameters have unique signatures for each category, allowing the RAG-enhanced LLM to conduct a step-by-step CoT reasoning process. The final recognition results are accurate and consistent with the actual materials, validating LLMaterial and confirming that it is indeed feasible to apply LLMs to identify materials from raw radar signals. 

\begin{table*}
  \caption{Ablation study.}
  \label{tab2}
  \resizebox{\textwidth}{!}{
  \begin{tabular}{c|ccccccc}
    \toprule
    Method& Metal bottle & Metal box & Glass bottle &Glass cup & Ceramic cup & Ceramic mug  & Plastic bottle\\
    \midrule
     LLMaterial w/o RAG&$\times$&$\surd$&$\times$&$\times$&$\times$&$\times$&$\surd$\\
    \rowcolor{gray!20} LLMaterial w RAG & $\surd$ &$\surd$ &$\surd$ &$\surd$ &$\surd$ &$\surd$ &$\surd$\\
  \bottomrule
\end{tabular}
}
\end{table*}

{\textbf{Analysis}}: To validate the effectiveness of our proposed LLMaterial, we performed an ablation study. As detailed in \Cref{tab2}, we compared our full system (LLMaterial w/ RAG) against an ablated version without this module (LLMaterial w/o RAG).
The results reveal a dramatic performance gap. The ablated version (LLMaterial w/o RAG) struggled in correctly identifying only two of the seven objects. In contrast, the LLMaterial system, empowered by RAG, achieved a perfect score, correctly identifying all seven objects. This demonstrates that the RAG module is not only beneficial but also indispensable, as it provides the essential context for the LLM to accurately interpret radar signals, thus enabling precise identification of object materials.

\section{Conclusion and future work}
In conclusion, this paper introduces LLMaterial, the first framework to successfully leverage LLMs for radar-based material identification. We tackle this task through two key innovations: (1) a physics-informed signal processing pipeline that transforms raw radar data into meaningful electromagnetic parameters, and (2) a RAG-enhanced LLM that interprets these parameters to infer the object's material composition. Our preliminary experiments confirm that LLMaterial achieves accurate material recognition, demonstrating for the first time that LLMs can effectively interpret object materials from raw radar signals.

We acknowledge several limitations in the current study. This work serves as a preliminary validation, focusing on only four material types. Future research should pursue three key directions: (1) broadening the diversity of evaluated materials to verify generalizability; (2) enhancing the RAG knowledge base and incorporating advanced CoT reasoning strategies to boost interpretability and accuracy; and (3) developing robust radar-vision fusion techniques for multimodal material recognition in real-world scenarios.


\section{Acknowledgments}
The work of He Chen is supported in part by the CUHK Strategic Seed Funding for Collaborative Research Scheme under Project 3136053 and the CUHK Direct Grant for Research under Project 4055229.

The research work described in this paper was conducted in the JC STEM Lab of Advanced Wireless Networks for Mission-Critical Automation and Intelligence funded by The Hong Kong Jockey Club Charities Trust.


\clearpage
\bibliographystyle{ACM-Reference-Format}
\bibliography{sample-base}

\end{document}